\documentclass[11pt,a4paper]{article}
\pdfoutput=1
\usepackage[utf8]{inputenc}
\usepackage{authblk}

\title{Quantum Prisoner's Dilemma and High Frequency Trading on the Quantum Cloud}

\author[1]{Faisal Shah Khan\thanks{Corresponding Author: quantumsheikh@gmail.com}}
\author[2]{Ning Bao\thanks{ningbao75@gmail.com}}
\affil[1]{Dark Star Quantum Lab, 211 Attain St, Fuquay-Varina, NC 27526}
\affil[2]{Computational Science Initiative, Brookhaven National Lab, Upton, NY, 11973}
\date{\today}

\usepackage{graphicx}
\usepackage{amsmath}
\usepackage{url}

\begin{document}

\maketitle

\begin{abstract}
High-frequency trading (HFT) offers an excellent user case and a potential killer application of the commercially available, first generation quasi-quantum communication and computation technologies. To this end, we offer here a simple but complete game-theoretic model of HFT as the famous two player game, Prisoner's Dilemma. We explore the implementation of HFT as a game on the (quasi) quantum cloud using the Eisert, Wilkens, and Lewenstein quantum mediated communication protocol, and how this implementation can increase transaction speed and improve the lot of the players in HFT. Using cooperative game-theoretic reasoning, we also note that in the near future when the internet is properly quantum, players will be able to achieve Pareto-optimality in HFT as an instance of reinforced learning.
\end{abstract}
\section{Introduction}

Non-cooperative game theory is the art of strategic interaction between individuals competing for joint stakes over which they entertain differing preferences. Game-theoretic reasoning can formally be traced back the Ancient Chinese General Sun Tzu (circa 500 BCE) and the ancient Indian minister, Chanakya (circa 300 BCE).

Mathematical formalization of non-cooperative game theory in the 20th century goes back to the work of von Neumann and Nash. The publication the seminal work of von Neumann and Morgenstern titled {\it Theory of Games and Economic Behavior} \cite{10.2307/j.ctt1r2gkx} brought focus upon game theory as the right mathematical language to analyze economic behavior and strategic decision making. The practical usefulness of the subject was made apparent by the awarding of several Noble prizes in Economics to developers of game-theoretic reasoning, including Nash \cite{Nash48}, Harsanyi \cite{Harsanyi}, Selten \cite{Selten1994}, Aumann \cite{Aumann}, and Smith \cite{Smith} for work in applications of game-theoretic reasoning to economics, political stratagem, and evolutionary biology. With the ongoing Covid-19 pandemic, game theoretic reasoning has also been used to shed light on best practices in developing optimal public health policy \cite{Elgazzar}. 

With the recent advent of commercially viable quantum computation and communication technologies, the confluence of ideas from game theory and quantum information processing has gained strong interest. This interest has given birth to the subject known as {\it quantum game theory} (section \ref{QG}), where the impact of quantum information technology on game-theoretic reasoning is studied. An area where quantum game theory may be of particular interest is the area of high-frequency trading. Here, many players participate in iterated buy/sell interactions at a very high rate, capitalizing on small market fluctuations in either duration, intensity, or both to gain revenue. Because the timing of such interactions is critical to the success of high-frequency firms compared to firms that trade at slower rates, new tools that can improve the degree of synchronicity between the firms and which can provide provably-secure communication, are of great interest. 

\section{Prisoner's Dilemma - A Game Theory Primer}

\begin{figure}
\centering
\includegraphics[scale=0.6]{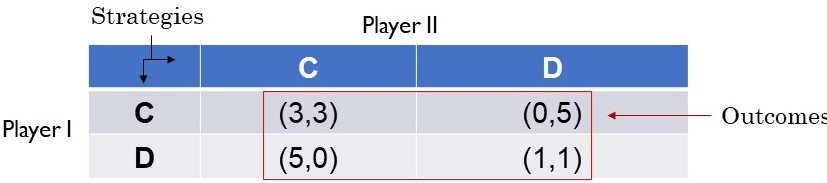}
\caption{Prisoner's Dilemma.}
\label{fig:PD1}
\end{figure}

Consider the non-cooperative game called {\it Prisoner's Dilemma}, a 2-player non-cooperative game in which each of the two players (prisoners) who committed a crime together are given the opportunity to reduce their time served in prison by helping authorities implicate the other player for the crime. This game is presented in tabular form in Figure \ref{fig:PD1} where the {\it outcomes} of the game are given as ordered pairs of numbers. The first number in each outcome is the {\it payoff} to Player I in the form of the number of years commuted from his sentence, and the second number is the payoff to Player II. 

The players have disparate preferences over the outcomes of the game, which are captured below using the symbol $\succ$ to denote the notion of ``preferred over'':
\begin{equation}\label{eq1}
\begin{split}
    &{\rm Player \hspace{2mm} I}: (5,0) \succ (3,3) \succ (1,1) \succ (0,5) \\
&{\rm Player \hspace{2mm} II}: (0,5) \succ (3,3) \succ (1,1) \succ (5,0).
\end{split}
\end{equation}

It is assumed that the players are {\it rational}, that is, each player will play the game in way that is consistent with his preferences. The game is played by employing {\it strategies} to optimize the payoffs. The two strategies available to both players are to either cooperate with the authorities to implicate the other player ({\bf C}), or to defect from offer to help the authorities ({\bf D}). The question is: what is the outcome of the game (or the play of the game)? 

The answer is provided in the form of {\it Nash equilibrium}, a profile of strategies, one per player, in which no player has motivation to deviate from his strategic choice. In other words, Nash equilibrium is a strategy profile in which each player's strategy is a best reply (with respect to the players' preferences) to all others. Not all games have a Nash equilibrium. 

\begin{figure}
\centering
\includegraphics[scale=0.6]{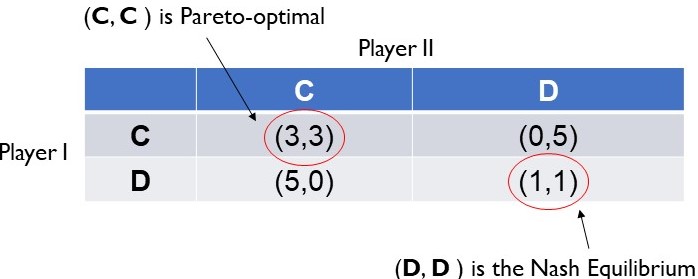}
\caption{Nash equilibrium versus Pareto-optimal outcomes in Prisoner's Dilemma.}
\label{fig:PD2}
\end{figure}

For Prisoners' Dilemma, Figure \ref{fig:PD2} shows that the Nash equilibrium is the strategy profile $({\bf D},\bf{D})$. This is the dilemma; for clearly, each player will be better off playing the strategy {\bf C}, but this is not a best reply to the strategic choice of {\bf C} by the other player. The strategy profile $({\bf C},\bf{C})$ (and its corresponding outcome) is {\it Pareto-optimal}, that is, its corresponding outcome is such that moving away from it to make one player better off will necessarily make another player worse off. Note that the strategy profiles $({\bf C},\bf{D})$ and $({\bf D},\bf{C})$ are also Pareto-optimal; however, no player wishes to complete her full sentence while her partner in crime walks free (as evidenced by the preference relations in expression (\ref{eq1})).

\subsection{Mixed Strategies and Mediated Communication }

When Nash equilibrium is not present in a game, or if it is sub-optimal, game-theorists suggest that players employ randomization over the outcomes as a mechanism for introducing or improving Nash equilibrium. To this end, players are allowed to independently randomize over their respective strategies, a notion referred to as {\it mixed strategies}, to produce probability distributions over the outcomes. The resulting {\it mixed game} will have at least one Nash equilibrium outcome (John Nash's Nobel prize winning result \cite{Nash48}). However, this mixed strategy Nash equilibrium need not be better paying than the one available in the original game, and it need not be Pareto-optimal. Indeed, this holds true for Prisoner's Dilemma. 

Further refinement of the Nash equilibrium may be possible if a referee is inducted into the game at negligible cost. This proper extension of a game is know as the {\it game with mediated communication}. In such games, the referee creates a probability distribution over the outcomes of the game that the players could not using mixed strategies. The referee then tells each player in confidence which strategy he should employ. Each player than checks the viability of the referee's advice with respect to his preferences and the 50-50 chance of the other player agreeing to the advice given to him by the referee. 
If the viability checks out, the player agrees with the referee. When both players agree to the referees advice, the resulting Nash equilibrium is known as a {\it correlated equilibrium}. 

Even further refinements of Nash equilibrium are conceivable by simply extending the domain of the game from Euclidean space to more exotic (and non-trivial) mathematical spaces such as Hilbert space and Banach space. The challenge then becomes how to keep the mathematical extensions grounded in physical reality. For the case of games extended to complex projective Hilbert space, the physical context is quantum mechanics. The result of this extension is the theory of ``quantum games''. 

\begin{figure}
\centering
\includegraphics[scale=0.7]{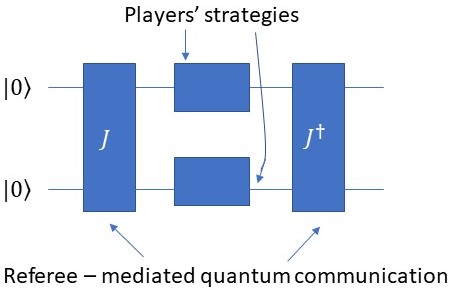}
\caption{The quantum circuit implementation of the EWL quantum game protocol. The referee consists of two quantum logic gates, $J$, which entangles the two qubits, and its inverse, $J^{\dag}$. In the middle of these two operations are the players' independent quantum strategic choices that each of them enacts on her qubit as unitary operations. We assume the top qubit is Player I's and the second one is Player II's.}
\label{fig:EWL}
\end{figure}

\section{Quantum Games}\label{QG}
Foreseeing the rise of quantum technologies like quantum computers and quantum communication devices and protocols, Meyer offered the first game-theoretic model of quantum algorithms. In his seminal work \cite{PhysRevLett.82.1052} on the topic, he showed that in a simple penny flipping game, the player with access to quantum physical operations (or ``quantum strategies'') acting on the penny always won the game. His work was followed by Eisert et al.'s work \cite{Eisert_1999} where the authors showed how to properly extend a game into the quantum realm with {\it quantum mediated communication}. These authors presented a two qubit (two player) quantum circuit that implemented the quantum communication protocol for Prisoner's Dilemma. This protocol is known as the EWL protocol and appears in figure \ref{fig:EWL}. 

The EWL protocol is a quantum circuit that takes in as input the two-qubit state 
\begin{equation}
    |00\rangle = \begin{pmatrix}
    1 \\
    0 \\ 
    0 \\ 
    0
    \end{pmatrix},
\end{equation}
with each qubit belonging to one player. This state is acted upon by the referee to produce a higher-order randomization in the form of a quantum superposition followed by measurement. In particular, the referee entangles the two qubits using a general entangling operator
\begin{equation}
    J(\gamma)=\cos \frac{\gamma}{2}I \otimes I + i \sin \frac{\gamma}{2}\sigma_x \otimes \sigma_x
\end{equation}
where $I$ is the $2 \times 2$ identity operator, $\sigma_x$ is the Pauli-spin flip operator, and $0 \leq \gamma \leq \frac{\pi}{2}$. When $\gamma=0$, the protocol reproduces the original ``classical'' game.

For $\gamma=\frac{\pi}{2}$, the game exhibits maximal entanglement between the qubits and the remarkable features discussed below. For this value of $\gamma$,  
\begin{equation}
 J =   \begin{pmatrix}
   \frac{1}{\sqrt{2}} & 0 & 0 & \frac{i}{\sqrt{2}}   \\
       0  & \frac{1}{\sqrt{2}} & \frac{i}{\sqrt{2}} & 0 \\
       0 & \frac{i}{\sqrt{2}} & \frac{1}{\sqrt{2}} & 0\\
       \frac{i}{\sqrt{2}} & 0 & 0  & \frac{1}{\sqrt{2}} 
    \end{pmatrix}.
\end{equation}
Therefore, 
\begin{equation}\label{ref}
    J |00 \rangle = \frac{1}{\sqrt{2}} \left( |00\rangle + i|11 \rangle \right)= \begin{pmatrix}
    \frac{1}{\sqrt{2}} \\
    0 \\ 
    0 \\ 
    \frac{i}{\sqrt{2}}
    \end{pmatrix}.
\end{equation}
The referee forwards the state in (\ref{ref}) to the players as her advice upon which the players can act with their quantum strategies. Finally, the referee disentangles the resulting two qubit state and makes a measurement, producing a probability distribution over the outcomes of the game (the observable states) from which expected payoffs to the players can be computed. Since the probability distribution was created using higher-order randomization by quantum superpositioning, the correlations it creates between the outcomes of the game after measurement are stronger than those possible classically \cite{Shimamura}. 

\subsection{(Almost) Solving the Dilemma}

The remarkable implication of the EWL protocol for Prisoner's Dilemma is that under the right subset of quantum strategies, this quantum extension of the game eliminates the dilemma and the resulting Nash equilibrium is Pareto-optimal! The quantum strategies that allow this are the two-parameter subset of the set of one qubit gates:
\begin{equation}\label{A}
  A:=  \left\{ \begin{pmatrix} 
    e^{i\phi}\cos \theta & \sin \theta \\
    -\sin \theta & e^{-i\phi}\cos \theta
    \end{pmatrix} : \hspace{1mm} 0 \leq \theta \leq \frac{\pi}{2}, \hspace{1mm} 0 \leq \phi \leq \frac{\pi}{2} \right\}.
\end{equation}

However, when the full set of quantum strategies is made available to the players \cite{FLITNEY2007381}, that is, 
\begin{equation}
   B:= \left\{ \begin{pmatrix}\label{B}
    e^{i\alpha}\cos \theta & e^{i\beta}\sin \theta \\
    -e^{-i\beta}\sin \theta & e^{-i\alpha}\cos \theta
    \end{pmatrix} : \hspace{1mm} 0 \leq \theta \leq \frac{\pi}{2}, \hspace{1mm}, \hspace{1mm} -\pi \leq \alpha, \beta \leq \pi \right\},
\end{equation}
the dilemma reappears in the quantum version of the game and the Nash equilibrium solution is the same as of the classical game. This is because a best reply to a quantum strategy from set $A$ is a quantum strategy from set $B$. But now, the other player also responds with a quantum strategy from set $B$, thus nullifying the quantum solution to the dilemma. 

Emulating mixed strategies, a further natural quantum extension is possible by allowing players to randomize over their quantum strategies, giving rise to the notion of {\it mixed quantum strategies}. Eisert et al. showed that while the players cannot solve the dilemma by resorting to mixed quantum strategies in Prisoner's Dilemma, they can come close to it. By using mixed quantum strategies, the players can affect a Nash equilibrium in which the payoff is $\left(2.5, 2.5 \right)$. This solution is closer to the Pareto-optimal outcome $(3,3)$ than the sub-optimal outcome $(1,1)$. Mixed strategies have a realistic physical interpretation as the result of quantum strategies being transmitted over a noisy communication channel. 

Motivated by the results of the seminal works of Meyer and Eisert et al., quantum game theory has become a major area of research since the seminal papers of Meyer and Eisert et al. A relatively recent and comprehensive review of the subject can be found in \cite{Khan2018}. 

\section{The Dilemma in High Frequency Trading}

High-frequency trading (HFT) is defined by Gomber et al. in \cite{Gomber} as follows.
\begin{quote}
     HFT  relates  to  the  implementation  of proprietary trading strategies by technologically advanced market participants. ... HFT  enable  market  participants  to dramatically  speed  up  the  reception  of  market  data,  internal  calculation  procedures,  order submission and reception of execution confirmations. 
\end{quote}
Our aim here is to show that quantum computing via the cloud can be used to implement HFT as a quantum game. For this, first note that HFT is an instance of Prisoner's Dilemma where Player I and Player II represent the trading mindset of a market, buying and selling of commodities using the two strategies {\bf Buy} or {\bf Sell}. Assuming that in markets there is a preference toward being part of a mass-buy versus a mass-sell, we set the following preferences for the players over the four possible strategy profiles as reasonably reflecting the mood of any market,
\begin{equation}\label{eq2}
\begin{split}
    &{\rm Player \hspace{2mm} I}: {\rm (Sell,Buy) \succ (Buy,Buy) \succ (Sell,Sell) \succ (Buy,Sell)} \\
&{\rm Player \hspace{2mm} II}:{\rm  (Buy,Sell) \succ (Buy, Buy) \succ (Sell, Sell) \succ (Sell,Buy),}
\end{split}
\end{equation}
with a player most preferring to sell on his terms versus buying on the other payers terms. 

\begin{figure}
\centering
\includegraphics[scale=0.6]{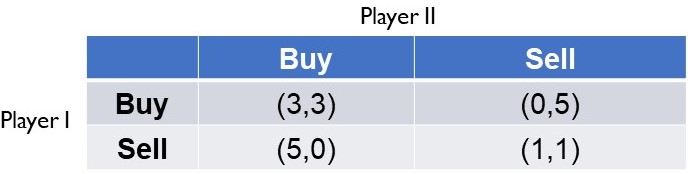}
\caption{High-frequency trading as an instance of Prisoners' Dilemma, as per the preferences described in expression (\ref{eq2}).}
\label{fig:HFTPD}
\end{figure}

These preferences are identical to those in Prisoners' Dilemma when the numerical payoff values from expression (\ref{eq1}) are faithfully substituted into expression (\ref{eq2}). Figure \ref{fig:HFTPD} shows HFT as an instance of Prisoners' Dilemma. Note that the dilemma in HFT is that the game will reach the sub-optimal Nash equilibrium $({\rm Sell}, {\rm Sell}) = (1,1)$, which is a highly detrimental outcome for markets. 
 
\subsection{HFT on the Quantum Cloud}

Today, the internet is quasi-quantum, meaning that users can access third party, first generation quantum processors via the cloud (the quantum cloud), which can offer transnational speed up. More importantly, the quasi-quantum internet can offer enhanced payoffs in the transaction when implemented using the EWL protocol for Prisoner's Dilemma.  

Due to the quasi-quantum nature of the internet, only noisy quantum communications are possible to date. Therefore, the referee will likely only be able to create limited entanglement between the qubits. This means that HFT on the quantum cloud will improve the lot of the players to only a near Pareto-optimal Nash equilibrium, the upper-limit of which for the moment is the appropriate equivalent of the notional $(2.5, 2.5)$ payoff. Nonetheless, even these small improvement in the payoffs will be worthwhile given the large amounts of money being traded. 

In the near future, the internet will be fully quantum, and improved fidelity of the transmission of the quantum information will mean that quantum entanglement between the players' qubits will be maintained for longer duration. This will allow the realization of the upper limit of the mixed quantum strategy Nash equilibrium, $(2.5, 2.5)$. 

\subsection{Optimality and cooperation in HFT on the quantum cloud}

From a non-cooperative game theory perspective, the pure quantum strategy Nash equilibrium that resolves the dilemma and produces the Pareto-optimal Nash equilibrium $(3,3)$ is fundamentally irrational. This is due to the fact that the best reply to any strategy from the set $A$ in (\ref{A}) is a strategy from the set $B$ in (\ref{B}). This would then seem to invalidate the whole idea of implementing HFT on the quantum internet of the near future for optimal benefits.
However, there is an appropriate game-theoretic solution for this issue found in the the cooperative theory of games. As Aumann points out in \cite{Aumann}: 
\begin{quote}
  We use the term cooperative to describe any possible outcome of a game, as long as no player can guarantee a better outcome for himself. It is important to emphasize that in general, a cooperative out-come is not in equilibrium; it’s the result of an agreement. For example, in the  well-known  “prisoner’s  dilemma”  game,  the  outcome  in  which  neither prisoner confesses is a cooperative outcome; it is in neither player’s best interests, though it is better for both than the unique equilibrium.
\end{quote}

Hence, the solution lies in the notion of agreement contracts and the ability to enforce them. For this, the game has to be played repeatedly and the behavioral history of the players collected and used to develop the contracts and the enforcement methods (incentives and disincentives). It is  noteworthy then that quantum games such as the quantum prisoner's dilemma can be thought of as the available policy space for an agent undergoing reinforcement learning. Here, however, it is known that the quantum policy options, in for example the quantum prisoner's dilemma, are Pareto-optimal over the classical policy options. Therefore, if the task undertaken in quantum reinforcement learning can be thought of as having instances of the prisoner's dilemma as subtasks, an agent with quantum strategies available to them will perform strictly better than one with only classical policy options, as observed by Meyer in his seminal work.

\section{Conclusion}

We established a game-theoretic interpretation of high-frequency trading  as the game Prisoner's Dilemma, and showed how it can be implemented as a quantum game using quantum computing processors available over the cloud. We argue that even today's nascent quantum technology infrastructure allows substantial improvement in the payoffs of the players of this game, and that in the near future, a fully quantum internet and better performing quantum processors will allow players to completely avoid the dilemma via reinforced learning of contracts, as predicted by cooperative game theory. 

\section*{Acknowledgments}
We would like to thank Nathan Benjamin, James Sully, Nathan Urban,  for useful discussions. N.B. is supported by the Computational Science Initiative at Brookhaven National Laboratory. 

\bibliographystyle{unsrt} 
\bibliography{references} 
 

 \end{document}